\documentclass[prb, preprint, floatfix]{revtex4-2}
\usepackage[utf8]{inputenc}
\usepackage{graphicx}
\usepackage{float}
\usepackage[version=4]{mhchem}
\usepackage{dcolumn}
\usepackage{bm}
\usepackage{units}
\usepackage{amsmath}
\usepackage{listings}
\usepackage[colorlinks=true, urlcolor=blue, linkcolor=black, citecolor=black]{hyperref}
\usepackage{soul}
\usepackage[dvipsnames]{xcolor}

\def\eV{\,\textrm{eV}}

\def\Ry{\,\textrm{Ry}}

\def\cm{\,\textrm{cm}}

\def\Ang{\,\textrm{\AA}}
\def\V{\,\textrm{V}}
\def\Acm2{\,\textrm{A}/\textrm{cm}^2}
\def\VA{\,\textrm{V}/\textrm{\AA}}
\def\eVA{\,\textrm{eV}/\textrm{\AA}}

\def\Vbot{V_\textrm{BG}}
\def\Vtop{V_\textrm{TG}}
\def\etal{\textit{et$\, $al.}}
\def\1st{1\textsuperscript{st}}
\def\2nd{2\textsuperscript{nd}}
\def\3rd{3\textsuperscript{rd}}
\def\4th{4\textsuperscript{th}}
\def\gates{Gate/[S-Mo-Se]\textsubscript{4}}
\def\gatese{Gate/[Se-Mo-S]\textsubscript{4}}
\def\hiji{\textit{h1}-\textit{h2}-\textit{h1}}
\def\hiii{\textit{h1}-\textit{h1}-\textit{h1}}
\def\hjjj{\textit{h2}-\textit{h2}-\textit{h2}}

\makeatletter
\renewcommand\@makecaption[2]{
  \par
  \vskip\abovecaptionskip
  \begingroup
  \small\rmfamily
    \begingroup
     \samepage
     \flushing
     \let\footnote\@footnotemark@gobble
     \@make@capt@title{#1}{#2}\par
    \endgroup
  \endgroup
  \vskip\belowcaptionskip
}
\makeatother

\begin{document}

\title{Multiple Control of Few-layer Janus MoSSe Systems} 

\def\kqz{
\author{Shuanglong Liu}
%\email{shlufl@ufl.edu}
\affiliation{Department of Physics, University of Florida, Gainesville, Florida 32611, USA}
\affiliation{Quantum Theory Project, University of Florida, Gainesville, Florida 32611, USA}

\author{James N. Fry}
%\email{fry@ufl.edu}
\affiliation{Department of Physics, University of Florida, Gainesville, Florida 32611, USA}

\author{Hai-Ping Cheng} 
\email{hping@ufl.edu} 
\affiliation{Department of Physics, University of Florida, Gainesville, Florida 32611, USA}
\affiliation{Quantum Theory Project, University of Florida, Gainesville, Florida 32611, USA}
\affiliation{Center for Molecular Magnetic Quantum Materials, University of Florida, Gainesville, Florida 32611, USA} 
}

%superscript affiliations
\author{Shuanglong Liu,$^{1,2}$}
%\email{shlufl@ufl.edu}
\author{James N. Fry,$^1$}
%\email{fry@ufl.edu}
\author{Hai-Ping Cheng$^{1,2,3}$}\email{hping@ufl.edu}

\affiliation{
$^1$Department of Physics, University of Florida, Gainesville, Florida 32611, USA \\
$^2$Quantum Theory Project, University of Florida, Gainesville, Florida 32611, USA \\
$^3$Center for Molecular Magnetic Quantum Materials, University of Florida, Gainesville, Florida 32611, USA} 

\date{\today}

\begin{abstract}
In this computational work based on density functional theory we study the electronic and electron transport properties of asymmetric multi-layer MoSSe junctions, known as Janus junctions. Focusing on 4-layer systems, we investigate the influence of electric field, electrostatic doping, strain, and interlayer stacking on the electronic structure. We discover that a metal to semiconductor transition can be induced by an out-of-plane electric field. The critical electric field for such a transition can be reduced by in-plane biaxial compressive strain. Due to an  intrinsic electric field, a 4-layer MoSSe can rectify out-of-plane electric current. The rectifying ratio reaches 34.1 in a model junction Zr/4-layer MoSSe/Zr. This ratio can be further enhanced by increasing the number of MoSSe layers. In addition, we show a drastic sudden vertical compression of 4-layer MoSSe due to in-plane biaxial tensile strain, indicating a second phase transition. Furthermore, an odd-even effect on electron transmission at the Fermi energy for Zr/$n$-layer MoSSe/Zr junctions with $n=1, \, 2,\, 3, \,\dots,\, 10$ is observed. These findings reveal the richness of physics in this asymmetric system and strongly suggest that the properties of 4-layer MoSSe are highly tunable, thus providing a guide to future experiments relating materials research and nanoelectronics. 

%\HPC{Do we need to perform DOS or orbital analysis? 
%Do we need to perform charge transfer analysis at interfaces?}

%\SLL{Shuanglong: A charge analysis is added in the end of the paper. A PDOS analysis is added in the section \ref{sec:efield}.}

\end{abstract}

%\pacs{Valid PACS appear here}

\maketitle

\section{Introduction} 
\label{sec:intro}

Two-dimensional (2D) transition metal dichalcogenides (TMDs) have 
potential applications in electronics/optoelectronics~\cite{RN170, RN376}, 
due to the presence of a direct band gap and sufficiently high mobility. 
The computed band gaps of TMDs vary by $\sim 2 \eV$, depending on
the chemical composition~\cite{RN381}, which provides rich opportunities
for different applications. 
In 2017, a semiconducting Janus transition metal dichalcogenide \ce{MoSSe} 
was synthesized in experiments independently by Zhang \etal~\cite{RN260}
and by Lu \etal~\cite{RN261}, where
One face of \ce{MoSSe} consists of S atoms and the other of \ce{Se} atoms. 
Such a discovery has inspired a surge of research into Janus 2D materials, 
including but not limited to \ce{MoSSe}~\cite{RN382, RN383, RN384, RN387, 
RN391, RN392, RN388, RN390, RN385, RN386, RN389}. 
For example, superior charge carrier mobility was predicted in 
monolayer \ce{WSSe}~\cite{RN382, RN383}, and 
large piezoelectricity was reported in monolayer \ce{Te2Se}~\cite{RN384}
and monolayer \ce{MoSO}~\cite{RN387}.
Strong Dzyaloshinskii-Moriya interaction was found in Janus manganese
dichalcogenides~\cite{RN391, RN392} and Janus chromium 
trihalides~\cite{RN388, RN390}. 

Since the electron affinity of S is lower than that of Se, there is electron
accumulation on the S side of monolayer \ce{MoSSe}.
Consequently, monolayer \ce{MoSSe} possesses an intrinsic out-of-plane 
electric dipole, pointing from S to Se. 
Multilayer \ce{MoSSe} can have a potential buildup in the out-of-plane
direction if the electric dipoles of individual layers are aligned in
the same direction. 
In this case, the band gap of multilayer \ce{MoSSe} decreases with the
number of \ce{MoSSe} layers $n$, until it closes at $n=4$~\cite{RN266}.
Since it is the intrinsic electric field that causes the band gap closing
in 4-layer \ce{MoSSe}, we wondered whether the band gap can reopen upon the 
application of a compensating external electric field. 
It is not a surprise, as we will show in the results section, that 
a metal to semiconductor transition in 4-layer \ce{MoSSe} can be induced by
an external out-of-plane electric field. 
However, the critical electric field that is required to induce the
metal to semiconductor transition in 4-layer \ce{MoSSe} is much smaller than
that for a semiconductor to metal transition in bilayer \ce{MoS2}~\cite{RN450}.
In the context of a field effect transistor, an external electric field can
be applied via a dual gate configuration, whereas a single gate 
configuration induces electrostatic doping. 
We will also show how the electronic structure of 4-layer \ce{MoSSe}
is affected by electrostatic doping. 
Furthermore, the internal electric field within 4-layer \ce{MoSSe} may
allow rectification of an out-of-plane electric current, which sets another
research goal of this study. 

Guo and Dong showed that the band gap of monolayer \ce{MoSSe} 
can be significantly tuned by in-plane biaxial strain.~\cite{RN397} 
Strain engineering of the electronic structure of heterogeneous bilayers
\ce{MoSSe}/\ce{WX_2} (X=S, Se)~\cite{RN393} and 
\ce{MoSSe}/\ce{WSSe}~\cite{RN394} have also been reported. 
These discoveries motivate us to investigate the influence of in-plane 
biaxial strain on the critical electric field for the metal to semiconductor
transition. 
In addition, we will also examine the influence of out-of-plane pressure
on the critical electric field. 

The rest of the paper is organized as follows. 
In Section \ref{sec:method}, we present the computational methods and
simulation parameters. 
We then show our computational results in Section \ref{sec:result},
which is further divided into subsections concerning the atomic 
structure, external electric field, electrostatic doping, strain, and 
rectifying effect respectively. 
Finally, conclusions are given in Section \ref{sec:conclusion}. 

\clearpage

\section{Method} 
\label{sec:method} 

Our calculations are based on density functional theory 
(DFT)~\cite{RN74, RN75} as implemented in the SIESTA 
package~\cite{RN267}. 
The effective screening medium (ESM) method is used to simulate
the effects of out-of-plane electric field and electrostatic 
doping~\cite{RN92, RN372}.
The $\textrm{DFT} + \textrm{NEGF}$ method~\cite{RN268,RN269, RN270} 
is used to simulate electron transport properties of
a Zr/$4$-layer \ce{MoSSe}/Zr junction. 
We calculate the electron transmission and the electric current via
the Caroli formula~\cite{RN275} and the Landauer 
formula~\cite{RN274,RN298} respectively. 

We apply a double-$\zeta$ polarized (DZP) basis set~\cite{RN267} 
to expand the Kohn-Sham orbitals and the electron density. 
A mesh cutoff of $150 \Ry$ is set to sample real space. 
A $15 \times 15$ Monkhorst-Pack $k$-point mesh~\cite{RN273} 
is chosen to sample the 2D reciprocal space for ionic relaxations. 
For self-consistent calculations, the $k$-point mesh is increased to
$21 \times 21$ to guarantee convergence. 
The $k$-point mesh is further increased to $101 \times 101$ for
calculating electron transmission. 
We adopt norm-conserving pseudo-potentials as generated by the 
Troullier-Martins scheme~\cite{RN271} and the Perdew-Burke-Ernzerhof
(PBE) exchange correlation energy functional~\cite{RN78}. 
The Van der Waals interaction is taken into account 
via the DFT-D2 method~\cite{RN272}.
The atomic structure of 4-layer \ce{MoSSe} is fully optimized in all 
our calculations except for those under finite bias, since the atomic
force under non-equilibrium conditions may not be reliable. 
The numerical tolerances for the density matrix, 
the Hamiltonian matrix, and the atomic force are no larger than
$1 \times 10^{-4}$, $2 \times 10^{-3} \eV$, and $0.02 \eVA$ 
respectively.

\section{Results} 
\label{sec:result} 

We present our results in five parts. 
In Section \ref{sec:pos}, we first display our atomic structure 
of 4-layer \ce{MoSSe}. 
Next, in Section \ref{sec:efield} we show 
a metal to semiconductor transition in 4-layer \ce{MoSSe}
induced by an out-of-plane external electric field. 
Third is Section \ref{sec:doping}, on the effects of electrostatic doping
due to a single back gate. 
Fourth, we examine how the critical electric field for the metal to semiconductor
transition is affected by in-plane biaxial strain in Section \ref{sec:strain}. 
Last, we demonstrate that 4-layer \ce{MoSSe} can rectify out-of-plane 
electric current in Section \ref{sec:transport}. 

\subsection{Atomic structure} 
\label{sec:pos}

\begin{figure}[htb!]
\centering
\includegraphics[width=8.5cm]{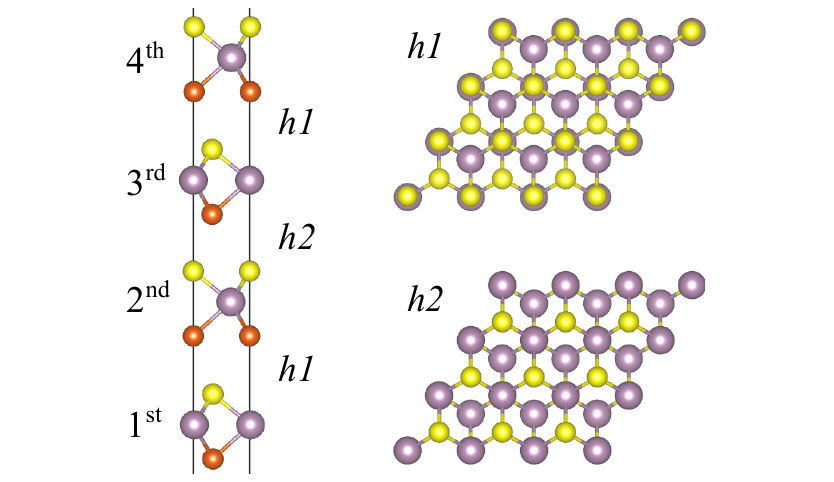} 
\caption{\label{fig:pos} 
Left: Side view of the 4-layer \ce{MoSSe} under study. 
%
% The four \ce{MoSSe} layers from the bottom to the top are, respectively,
% labeled as the \1st{}, \2nd{}, \3rd{}, and \4th{} layer. 
%
Label \textit{h1} (\textit{h2}) identifies a type 1 (type 2) hollow site interlayer stacking. 
Upper (lower) right: Top view of a bilayer \ce{MoSSe} with \textit{h1}
(\textit{h2}) interlayer stacking. 
} 
\end{figure} 

Our atomic structure of 4-layer \ce{MoSSe} is shown on the left of
Fig.~\ref{fig:pos}. 
The S (Se) side of each \ce{MoSSe} layer faces upward. 
Adjacent \ce{MoSSe} layers adopt a hollow site interlayer stacking, 
which is energetically more favorable than on-top site interlayer
stacking~\cite{RN373, RN374}.
There are two types of hollow site interlayer stacking, 
as shown on the right in Fig.~\ref{fig:pos}. 
Type 1 hollow site interlayer stacking (\textit{h1}) is slightly higher 
in energy than type 2 hollow site interlayer stacking 
(\textit{h2})~\cite{RN373, RN374}.
In this study, we focus on the 4-layer \ce{MoSSe} with a stacking sequence
of \hiji{}. 
Nevertheless, we also verify our major findings for
\hiii{} stacking 4-layer \ce{MoSSe} and
\hjjj{} stacking 4-layer \ce{MoSSe}.

\subsection{Out-of-plane electric field} 
\label{sec:efield}

\begin{figure}[htb!]
\centering
\includegraphics[width=8.5cm]{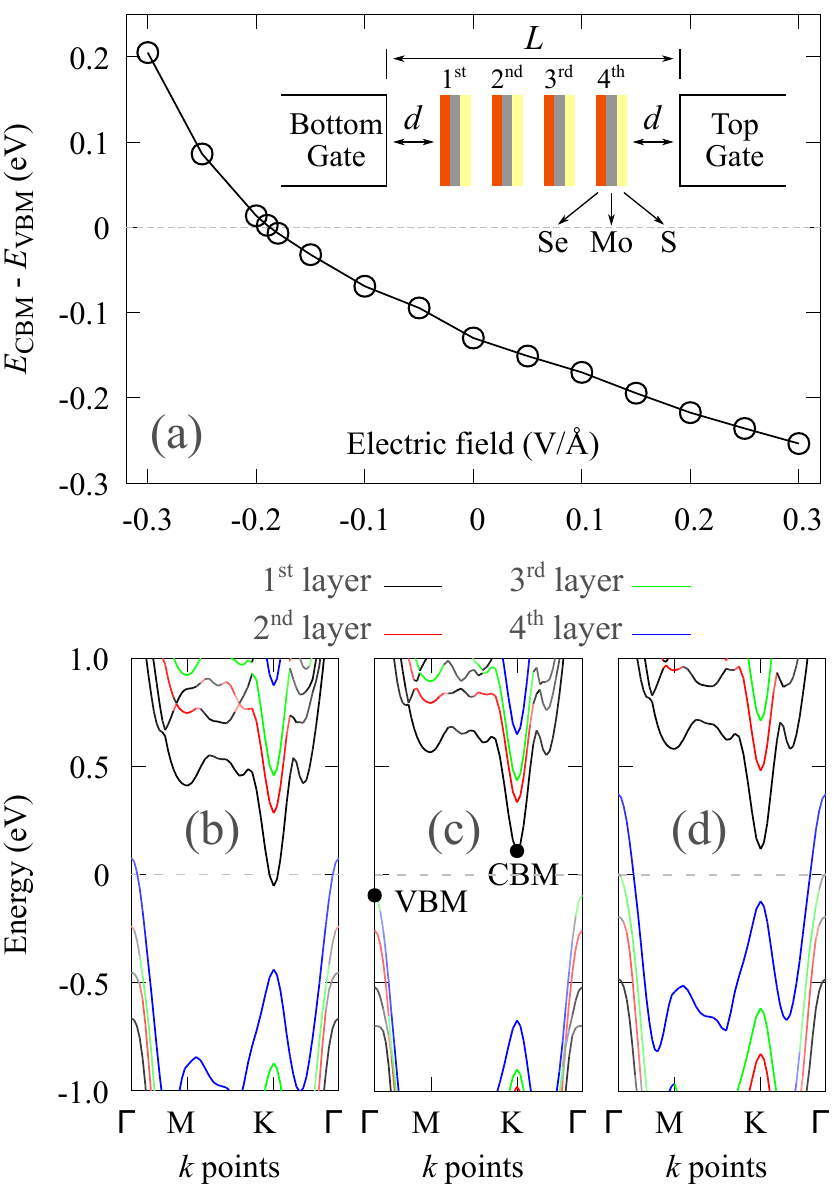} 
\caption{\label{fig:efield} 
(a) Energy difference between the conduction band minimum (CBM) 
and the valence band maximum (VBM) of the 4-layer \ce{MoSSe} 
under different electric fields. 
The inset illustrates the system setup.
(b)--(d) Band structures of the 4-layer \ce{MoSSe} under zero, 
$-0.30$, and $+0.30 \VA$ electric fields respectively. 
Colors identify the layer from which the band mainly originates 
(see text),
The CBM and the VBM are marked in panel (c). 
} 
\end{figure}

The inset of Fig.~\ref{fig:efield}a illustrates our model setup for applying 
an out-of-plane electric field to 4-layer \ce{MoSSe}. 
The 4-layer \ce{MoSSe} is placed between two metal plates, a bottom
gate and a top gate, separated by a distance $L$. 
The two gates have constant electrostatic potentials, $\Vbot$ ($\Vtop$)
for the bottom (top) gate.
Electric field in this work refers to the average electric field
between the two gates, 
\begin{equation}
\label{eqn:efield}
E = (\Vtop - \Vbot)/L. 
\end{equation} 
Each gate is spaced from the Janus \ce{MoSSe} by a distance 
$d \sim 15\Ang$ of vacuum.
The four \ce{MoSSe} layers from bottom to the top are labeled 
as \1st{}, \2nd{}, \3rd{}, and \4th{} layer respectively . 
The sulfur side faces the top gate. 

Fig.~\ref{fig:efield}b shows the band structure of the 4-layer \ce{MoSSe}
under zero electric field. 
Roughly speaking, an energy band is colored black (red, green, blue) 
if it originates mainly from the \1st{} (\2nd{}, \3rd{}, \4th{}) layer. 
The details of colormap are given in Appendix \ref{app:colormap}. 
We see from Fig.~\ref{fig:efield}b that the 4-layer \ce{MoSSe} is metallic with 
the \1st{} (\4th{}) layer being electron- (hole-) doped. 
If a negative electric field of $-0.30$ $\VA$ is applied, an energy gap of 
$0.20 \eV$ opens up around the Fermi energy, and the 4-layer \ce{MoSSe} 
becomes a semiconductor (Fig.~\ref{fig:efield}c). 
In contrast, the 4-layer \ce{MoSSe} remains a metal if a positive electric 
field of $+0.30$ $\VA$ is applied (Fig.~\ref{fig:efield}d). 
In order to find the critical electric field for the metal to semiconductor transition, 
we examine the energy difference between the conduction band minimum 
(CBM) and the valence band maximum (VBM) versus electric field in 
Fig.~\ref{fig:efield}a. 
The energy difference $E_\textrm{CBM} -  E_\textrm{VBM}$ decreases with 
electric field, and it reaches zero at a critical electric field of 
$\sim -0.187 \VA$.

% Note that the \1st{} \ce{MoSSe} layer loses electrons to the \4th{} 
% \ce{MoSSe} layer at $-0.30$ $\VA$ electric field, whereas both these 
% two layers lose electrons to the \3rd{} \ce{MoSSe} layer at $0.30$ $\VA$ 
% electric field. 
% %
% In the latter case, the conduction band of the \1st{} layer becomes 
% completely empty. 
% %
% Thus, the \3rd{} and \4th{} layers should be the only conducting parts of 
% the 4-layer \ce{MoSSe} at $0.30$ $\VA$ electric field at low temperatures. 
These results are for the 4-layer \ce{MoSSe} with 
\hiji{} stacking. 
The critical electric field for the \hiii{} and \hjjj{} 
stackings are $-0.252 \VA{}$ and $ -0.145 \VA{}$, respectively. 
Compared with the \hiji{} stacking, the
\hiii{} (\hjjj{})
stacking has a larger (smaller) critical field in magnitude. 
In order to understand such a difference, we analyse the internal potential
buildup between adjacent \ce{MoSSe} layers. 
In practice, we take the planar average of the electrostatic potential
over the 2D plane parallel to the Janus \ce{MoSSe}. 
Denoting the plane averaged electrostatic potential at the vertical position
of a Se/S atom as $V_\textrm{Se/S}$, 
we find that the potential buildup $V_\textrm{Se} - V_\textrm{S}$ at the three
van der Waals gaps of the 4-layer \ce{MoSSe} with 
\hiji{} stacking are $1.39 \eV$ (\textit{h1}), 
$1.22 \eV$ (\textit{h2}), and $1.44 \eV$ (\textit{h1}) respectively. 
As such, the potential buildup at a van der Waals gap between two $\ce{MoSSe}$
layers with \textit{h1} stacking is larger than that for \textit{h2} stacking by
$0.17/0.22 \eV$. 
A larger internal electric field requires a larger external electric field
to compensate and to open a band gap around the Fermi level. 
Therefore, the critical electric field for the \hiii{}
(\hjjj{}) stacking is larger (smaller) than that 
for the \hiji{} stacking.

\begin{figure}[htb!]
\centering
\includegraphics[width=8.5cm]{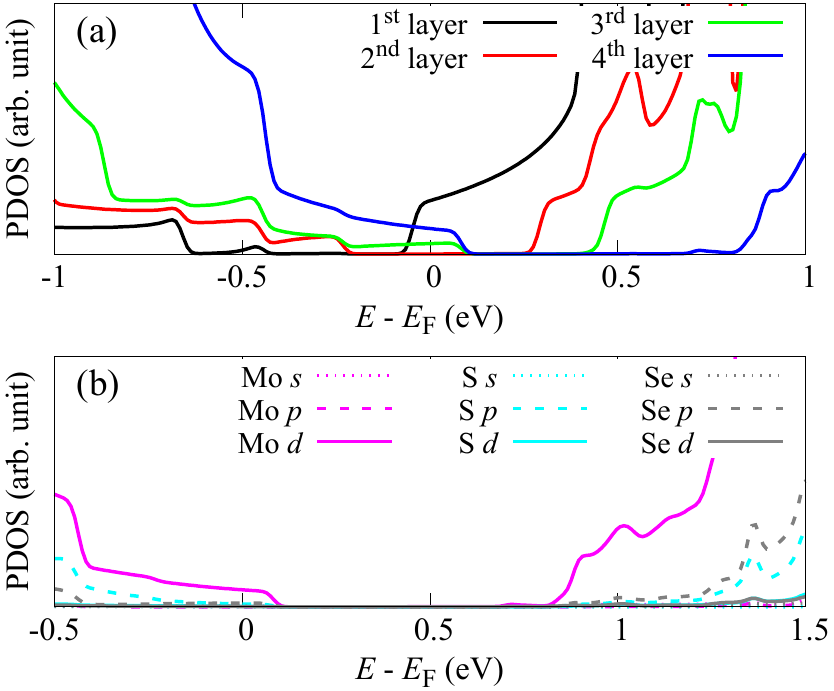} 
\caption{\label{fig:pdos} 
Projected density of states (PDOS) of 
(a) each \ce{MoSSe} layer, and 
(b) the $s$-, $p$-, and $d$-orbitals in the \1st{} \ce{MoSSe} layer. 
The Fermi energy is set to zero. 
} 
\end{figure}

Fig.~\ref{fig:pdos}a shows the layer-decomposed projected density of states (PDOS) 
for 4-layer \ce{MoSSe} with \hiji{} stacking under
zero external electric field.
Recall that 4-layer \ce{MoSSe} is metallic in such a way that both
the valence and the conduction bands cross the Fermi energy. 
As seen from Fig.~\ref{fig:pdos}a, the conduction-band states around
the Fermi level originate merely from the \1st{} \ce{MoSSe} layer while
the valence-band states around the Fermi level originate from both the \3rd{} and the \4th{} \ce{MoSSe} layers. 
Furthermore, the PDOS of the \4th{} \ce{MoSSe} layer is more than two times
larger than the PDOS of the \3rd{} \ce{MoSSe} layer at the Fermi level. 
There are not any states originating from the the \2nd{} \ce{MoSSe}
layer at the Fermi level. 
Fig.~\ref{fig:pdos}b shows the orbital-decomposed PDOS for the \4th{}
\ce{MoSSe} layer. 
The states around both the VBM and the CBM are dominated by \ce{Mo} $d$ orbitals. 
The contribution from S or Se $p$ orbitals to the valence (conduction)
band becomes larger at lower (higher) energies. 
The contributions from all other orbitals are small within the
energy range of $[-0.5:1.5] \eV{}$. 
%
% The orbital-decomposed PDOS of the other three \ce{MoSSe} layers are
% similar to that of the \4th{} \ce{MoSSe} layer despite of a shift in
% energy. 
%
An analysis of the orbital-decomposed PDOS for the \3rd{} \ce{MoSSe}
layer (not shown in the figure) shows that there is a significant contribution
of the S $p$ orbitals to the states at the Fermi level. 
This indicates a hybridization between the \3rd{} and the \4th{}
\ce{MoSSe} layers.

\subsection{Electrostatic doping}
\label{sec:doping}

\begin{figure}[htb!]
\centering
\includegraphics[width=8.5cm]{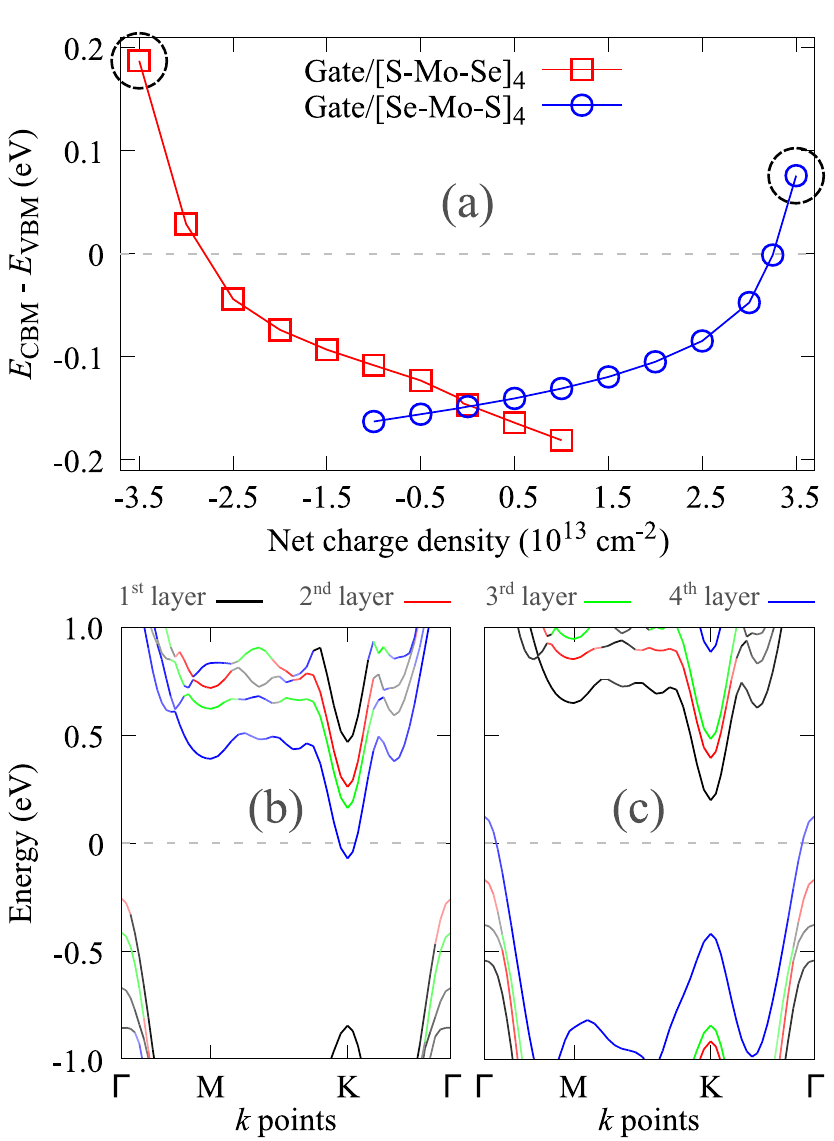} 
\caption{\label{fig:doping} 
(a) Energy difference between conduction band minimum (CBM) and 
valence band maximum (VBM). 
For red squares (blue circles) the S (Se) side of 4-layer \ce{MoSSe} faces
the back gate. 
(b) [(c)] Energy bands of \gates{} [\gatese{}] under a doping level of 
$-3.5$ [$+3.5$] $\times 10^{13} \cm^{-2}$ as 
indicated by the dashed circles in panel (a). 
Counting from the back gate, the four \ce{MoSSe} layers are referred to
as the \1st{}, \2nd{}, \3rd{}. and \4th{} layer respectively.
} 
\end{figure} 

In the previous section, we have shown that a metal to semiconductor transition
can be induced by an out-of-plane electric field when the system is charge
neutral. 
In this section, we examine the effects of electrostatic doping (charging)
due to a back gate. 

Fig.~\ref{fig:doping}a shows the energy difference between the CBM
and the VBM versus doping level for the 4-layer \ce{MoSSe} with \hiji{}
stacking subject to a single back gate. 
An energy gap between the CBM and the VBM can be induced by 
electron (hole) doping if the S (Se) side of 4-layer \ce{MoSSe} faces
the back gate. 
The gap-inducing external electric field for both \gates{} (the S side 
faces the back gate) and \gatese{} (the Se side faces the back gate)
is from the S side to the Se side of the 4-layer \ce{MoSSe}, 
in accordance with the results in the previous section. 
Fig.~\ref{fig:doping}b (Fig.~\ref{fig:doping}c) shows the band structure
of \gates{} (\gatese{}) with a doping level such that an energy gap 
between the CBM and the VBM occurs. 
As seen from the figures, the Fermi level still crosses the conduction
(valence) band. 
Therefore, electrostatic doping due to a single back gate does not induce
metal to semiconductor transition. 

Figs.~\ref{fig:doping}b and \ref{fig:doping}c also show that only the
\4th{} \ce{MoSSe} layer is conducting at low temperatures after an 
energy gap opens.
Note that the \4th{} \ce{MoSSe} layer is farthest away from the back gate. 
This phenomenon may be useful for electronic devices in which delicate 
control of conducting channels is required.

\subsection{In-plane biaxial strain}
\label{sec:strain}

\begin{figure}[htb!]
\centering
\includegraphics[width=8.5cm]{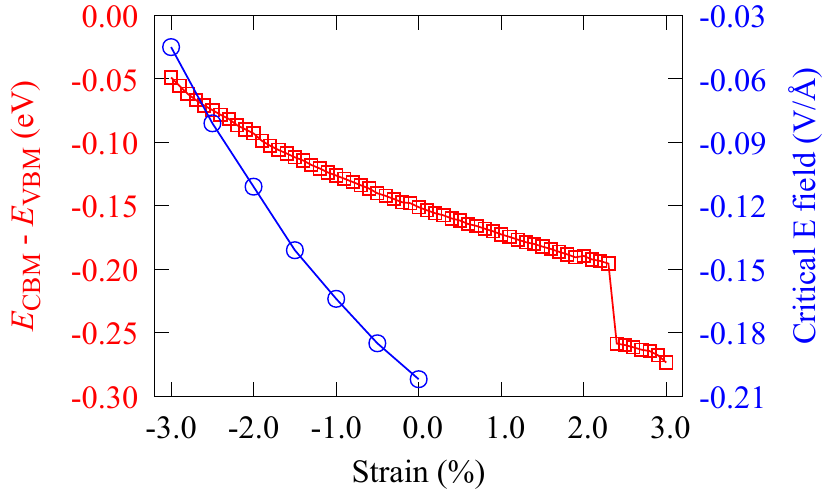} 
\caption{\label{fig:strain} 
Red squares show the energy difference between the conduction band minimum (CBM) and the valence band maximum (VBM) for 4-layer \ce{MoSSe} as a function of in-plane biaxial strain (left-hand scale)
Blue circles show the critical electric field for metal to semiconductor transition versus strain (right-hand scale).
} 
\end{figure} 

The critical electric field of $-0.187 \VA$ is gigantic. 
In this section, we examine whether the critical field for the metal to semiconductor 
transition can be reduced by in-plane biaxial strain. 

The red squares in Fig.~\ref{fig:strain} shows the energy difference
$E_\textrm{CBM} -  E_\textrm{VBM}$ versus strain. 
The strain is measured by $(a-a_0)/a_0$, where $a$ is the lattice constant under a finite strain and $a_0$ is the lattice constant under zero strain. 
A positive (negative) strain indicates tension (compression). 
$E_\textrm{CBM} -  E_\textrm{VBM}$, which is always negative here, decreases (increases) with tensile (compressive) strain. 
Intuitively, a smaller $\lvert E_\textrm{CBM} -  E_\textrm{VBM} \rvert$ would
require a smaller electric field to open an energy gap between the CBM and the VBM. 
Therefore, we search for the critical electric field at several compressive strains, and show the results as the blue circles in Fig.~\ref{fig:strain}. 
Indeed, the magnitude of the critical field decreases with compressive strain. 
Under a compressive strain of $-3.0\%$, the magnitude of the critical field is reduced to $-0.045 \VA$, which is more feasible in experiments. 

For the 4-layer \ce{MoSSe} with \hiii{} stacking under a compressive
strain of $-3.0\%$, $E_\textrm{CBM} -  E_\textrm{VBM}$ becomes $-0.062 \eV$. 
The corresponding critical electric field is $-0.112 \VA$ which is 
also significantly smaller in magnitude than the zero-strain value
(of $-0.252 \VA{}$) for the same stacking. 
$E_\textrm{CBM} -  E_\textrm{VBM}$ is approximately $0.007 \eV$ for the
4-layer \ce{MoSSe} with \hjjj{} stacking under a compressive strain of
$-3.0\%$. 
A metal to semiconductor transition can be induced by a moderate
compressive strain alone for the \hjjj{} stacking. 
The critical in-plane biaxial strain is about $-2.9\%$. 

At a strain of $\sim 2.3\%$, the energy difference 
$E_\textrm{CBM} -  E_\textrm{VBM}$ exhibits a sudden decrease. 
This is accompanied with a sharp change in the thickness of the 4-layer
\ce{MoSSe}, by about $0.47 \Ang{}$. 
More structural details concerning this sharp change are given in Appendix \ref{sec:astruct}.
It is worth mentioning that the function of energy versus strain is smooth
even at the strain of $\sim 2.3\%$. 

We additionally simulate out-of-plane pressure by reducing the thickness of the 4-layer \ce{MoSSe} with \hiji{} stacking. 
All atoms are relaxed except the bottom-most atomic layer and the 
top-most atomic layer. 
Our simulations show that $E_\textrm{CBM} -  E_\textrm{VBM}$ decreases
(becoming more negative) as the thickness is reduced. 
For example, a reduction in the thickness by $0.5 \Ang$ leads to a 
decreases in $E_\textrm{CBM} -  E_\textrm{VBM}$ by about $0.06 \eV$. 
Thus, the critical electric field for metal to semiconductor transition is likely to increase with out-of-plane pressure.

\subsection{Rectifying effect}
\label{sec:transport} 

\begin{figure}[htb!]
\centering
\includegraphics[width=8.5cm]{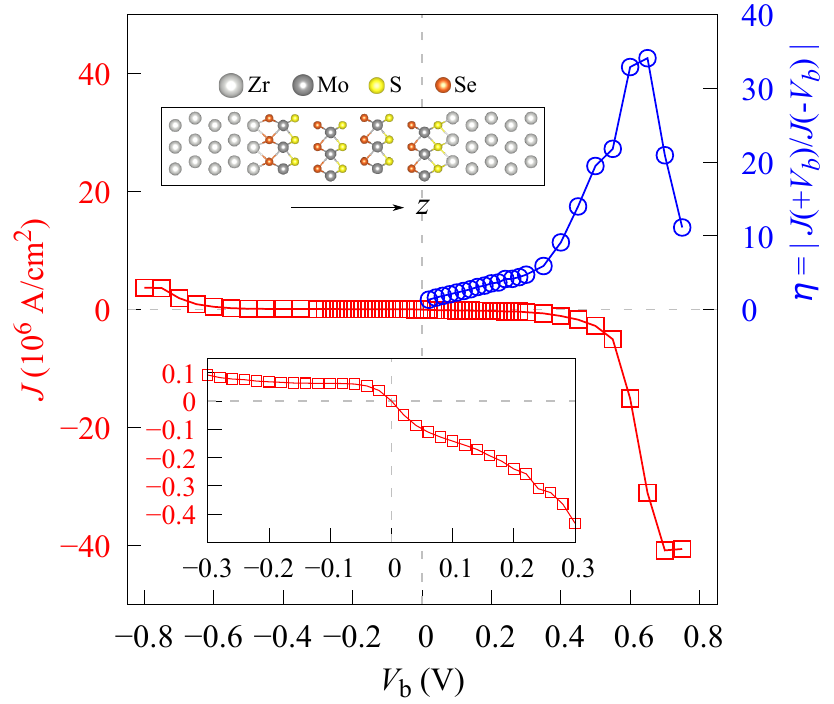} 
\caption{\label{fig:iv} 
Red squares show the electric current density in a Zr/4-layer \ce{MoSSe}/Zr junction as a function of bias voltage. 
Blue circles show the rectifying ratio of the junction versus bias voltage. 
The inset zooms in on the central part of the junction. 
} 
\end{figure} 

Since 4-layer \ce{MoSSe} has a substantial internal electric field, it may 
rectify out-of-plane electric current when employed in a circuit. 
Here, we present a proof-of-concept calculation where the 4-layer
\ce{MoSSe} with \hiji{} stacking is sandwiched between two zirconium
electrodes, as shown in the upper inset of Fig.~\ref{fig:iv}. 
A Zr electrode is chosen due to a small lattice mismatch with \ce{MoSSe}. 
The Zr/4-layer \ce{MoSS}/Zr junction is periodic in the $x$- and 
$y$-directions, 
and electron transport is along the $z$-direction. 
Although only two finite pieces of Zr are shown in the inset, 
the left and right electrodes extend to infinity.

The red squares in Fig.~\ref{fig:iv} show the calculated electric current density $J$ in the Zr/4-layer \ce{MoSS}/Zr junction as a function of bias voltage. 
The bias voltage $V_b$ is defined by 
\begin{equation}
V_b = \mu_L - \mu_R,
\end{equation}
where $\mu_L$ ($\mu_R$) is the chemical potential of the left (right)
electrode. 
The $J$-$V_b$ curve at low bias voltages is zoomed in and shown in the 
lower inset. 
$J(-0.02 \V)$ is about $3.7 \times 10^{4} \Acm2$ which is on the same 
order with $J(+0.02 \V) \sim -5.0  \times 10^{4} \Acm2$. 
Defining a rectifying ratio $\eta$ as 
\begin{equation}
\eta = \left \lvert J(+V_b)/J(-V_b) \right\rvert, 
\end{equation}
we see that $\eta(0.02 \V)$ is as small as 1.4. 
As shown by the blue circles in Fig.~\ref{fig:iv}, $\eta$ first gradually increases with bias voltage, reaching 4.7 at $0.3 \V$. 
As the bias voltage further increases in the positive direction, 
a sharp increase in the electric current density occurs at around $0.6 \V$. 
Such a sharp increase is however absent at negative bias voltages. 
As a result, a rectifying ratio as high as 34.1 is achieved at $0.65 \V$. 
$\eta$ begins to decrease after $0.65 \V$, and is reduced to 11.1 
at $0.75 \V$. 

\begin{figure}[htb!]
\centering
\includegraphics[width=8.5cm]{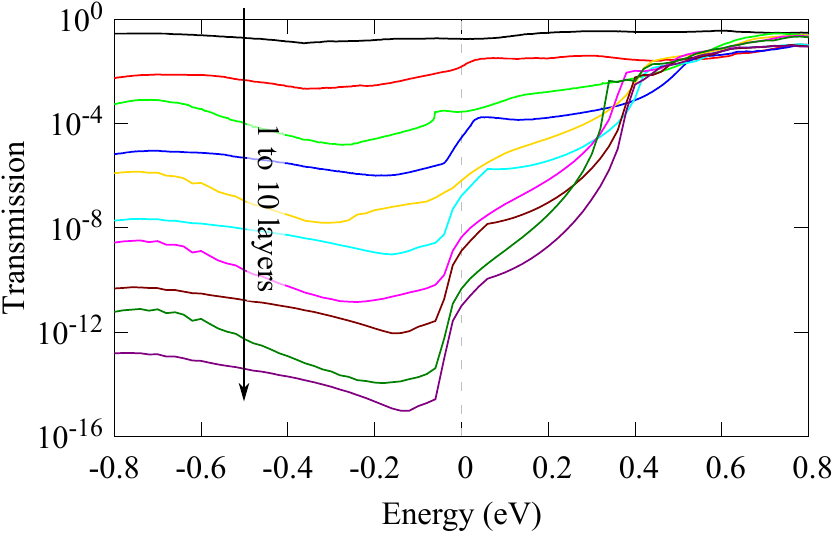} 
\caption{\label{fig:transmission} 
Electron transmission versus energy for Zr/$n$-layer \ce{MoSSe}/Zr junctions 
with $n = 1, 2, 3, \dots, 10$. 
The Fermi energy is set to zero. 
} 
\end{figure}

We next examine whether the rectifying ratio can be enhanced by tuning
the number of \ce{MoSSe} layers. 
Fig.~\ref{fig:transmission} shows electron transmission as a function
of energy $T_n(E)$ for Zr/$n$-layer \ce{MoSSe}/Zr junctions under zero
bias with $n = 1, 2, 3, \dots, 10$. 
%
%For all these junctions, the electron transmission at positive energies 
%(above Fermi energy) is larger than that at negative energies (below Fermi
%energy). 
%
Setting the Fermi energy to zero, we observe that $T_n(E)$ decays exponentially with $n$ for $E \in [-0.8: 0] \eV$. 
Particularly, $T_{10}(E)$ is more than 10 orders of magnitude smaller 
than $T_{1}(E)$ within this energy range. 
In contrast, $T_n(E)$ is always above $10^{-2}$ (and smaller than 1) 
regardless of $n$ for $E \in [0.5:0.8] \eV$. 
Thus, we infer from the zero bias transmission function that thicker \ce{MoSSe}
should exhibit a higher rectifying ratio. 
We verify this idea using a Zr/5-layer \ce{MoSSe}/Zr junction and obtain the higher value $\eta = 85.6$ at $V_b = 5.5 \V$. 

The electron transmission at Fermi energy decays differently with the number of \ce{MoSSe} layers $n$  depending on the parity of $n$. 
This odd-even effect originates from the parity-dependent potential buildup
at the contact between the $n$-layer \ce{MoSSe} and the right Zr electrode. 
Further analysis of this phenomenon is presented in Appendix \ref{sec:parity}.

\section{Conclusion} 
\label{sec:conclusion}

We have performed a detailed study of the electronic and electron transport properties of $n$-layer
\ce{MoSSe} ($n=1-10$) with focus on $n=4$ using first principles based density functional theory. Several significant findings obtained are as follows:
1) We show that a metal to semiconductor transition can be induced by 
an out-of-plane electric field; 
2) Sensitive to the stacking order, the critical electric fields for the 4-layer \ce{MoSSe} with \hiii{},
\hiji{}, and \hjjj{} stackings are, respectively, $-0.252$, $-0.187$, and 
$-0.145 \VA$;
3) The critical electric field for the \hiii{} stacking is highest
(in magnitude) because the internal potential buildup with a \textit{h1}
stacking is larger than that with a \textit{h1} stacking.
We can reduce the critical electric field by applying a in-plane biaxial
compressive strain, and this works for all the three stackings;
4) The applied strain induces a second phase transition indicated by a sudden compression in the vertical dimension;
5) The 4-layer \ce{MoSSe} with \hiji{} stacking is always metallic
upon electrostatic doping, although an energy gap between the conduction minimum and the valence maximum can be opened; 
6) When sandwiched between two Zr electrodes, the 4-layer \ce{MoSSe}
can rectify electric current with a maximum rectifying ratio of 34.1 
at $0.65 \V$; and 
7) finally, the rectifying ratio can be enhanced by increasing the number
of \ce{MoSSe} layers. 
Concluded from a high-level computational approach, these results have predicting power. One can use a combination of these controlling parameters to guide future experiments in areas of material design and nanoelectronics.

\begin{acknowledgments}

This work is supported by the US Department of Energy (DOE), Office of Basic Energy Sciences (BES), under Contract No.~DE-FG02-02ER45995. Computations were done using the utilities of the National Energy Research Scientific Computing Center and University of Florida Research Computing.

\end{acknowledgments}

\appendix

\renewcommand{\thefigure}{\thesection\arabic{figure}}

\section{Colormap for Fig.~\ref{fig:efield}}
\label{app:colormap}

\setcounter{figure}{0}  

\begin{figure}[htb!]
\centering
\includegraphics[width=3.6cm]{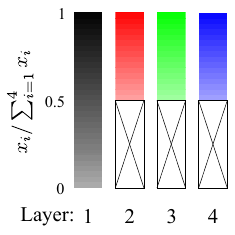} 
\caption{\label{fig:colormap} 
Colormap of the relative projected density of states (PDOS)
from each \ce{MoSSe} layer. 
$x_i$ represents the PDOS from the $i$th layer. 
} 
\end{figure} 

Fig.~\ref{fig:colormap} shows the colormap for Fig.~\ref{fig:efield}
in the main text. 
It is generated using the following script for Gnuplot.
\begin{lstlisting}[basicstyle=\footnotesize]
set palette model RGB 
set palette color
set palette defined ( \
0.0 "#aaaaaa", 0.2 "#808080", 1.0 "#000000", \
1.1 "#ffffff", 1.2 "#ffffff", 2.0 "#ff0000", \
2.1 "#ffffff", 2.2 "#ffffff", 3.0 "#00ff00", \
3.1 "#ffffff", 3.2 "#ffffff", 4.0 "#0000ff" )

$g(x_1, x_2, x_3, x_4)$ = \
$x_1$/($x_1$+$x_2$+$x_3$+$x_4$) > 0.5 ? $x_1$/($x_1$+$x_2$+$x_3$+$x_4$) : \ 
$x_2$/($x_1$+$x_2$+$x_3$+$x_4$) > 0.5 ? $x_2$/($x_1$+$x_2$+$x_3$+$x_4$) + 1 : \ 
$x_3$/($x_1$+$x_2$+$x_3$+$x_4$) > 0.5 ? $x_3$/($x_1$+$x_2$+$x_3$+$x_4$) + 2 : \ 
$x_4$/($x_1$+$x_2$+$x_3$+$x_4$) > 0.5 ? $x_4$/($x_1$+$x_2$+$x_3$+$x_4$) + 3 : \
$x_1$/($x_1$+$x_2$+$x_3$+$x_4$)

p "bands.dat" u 1:2:(g($\$3, \$4, \$5, \$6$)) w l palette
\end{lstlisting}
In this script, $x_i$ is the projected density of states (PDOS) 
of the $i$th \ce{MoSSe} layer due to a Kohn-Sham state.
If the ratio $\tau_i = x_i/(x_1+x_2+x_3+x_4)$ is greater than 0.5, 
then the contribution from the $i$th layer dominates. 
We identify the dominant layer index by color type (black, red, green,
or blue), and represent the value of $\tau_i$ by color brightness. 
If none of $\tau_i$ ($i=1,2,3,4$) is greater than 0.5, then we use
a grayscale according to the value of $\tau_1$.

\section{Abrupt structural change}
\label{sec:astruct}

\setcounter{figure}{0} 

\begin{figure}[htb!]
\centering
\includegraphics[width=8.5cm]{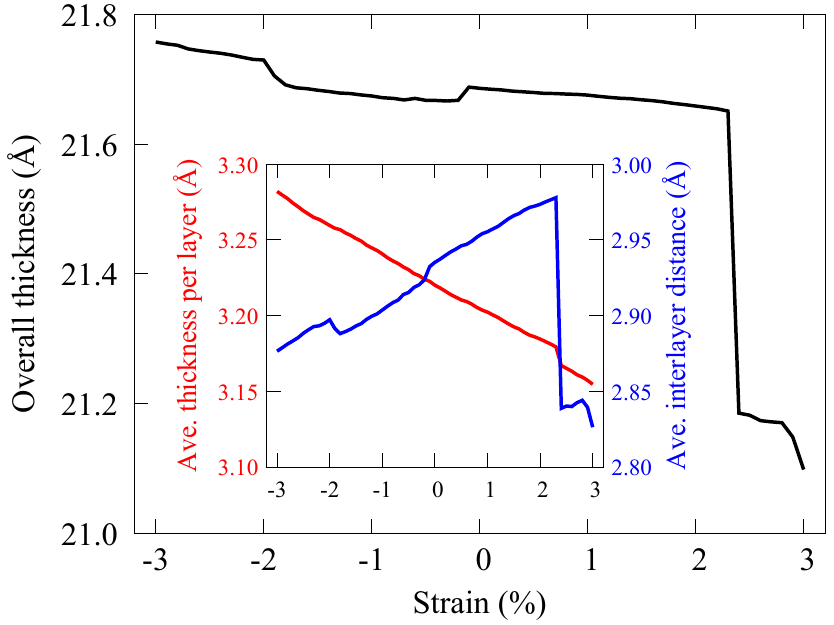} 
\caption{\label{fig:thickness} 
Thickness of 4-layer \ce{MoSSe} versus in-plane biaxial strain. 
Inset: Average thickness per \ce{MoSSe} layer and average interlayer
distance versus in-plane biaxial strain. 
} 
\end{figure} 

Fig.~\ref{fig:thickness} shows the overall thickness of 4-layer \ce{MoSSe}
with \hiji{} stacking versus in-plane biaxial strain. 
The thickness gradually decreases with tensile strain before a sharp decrease of $0.47 \Ang{}$ occurs at a strain of 2.3\%.
This sharp decrease is due to both an abrupt decrease in the thickness of an
individual \ce{MoSSe} layer and an abrupt decrease in the interlayer 
distance (see the inset of Fig.~\ref{fig:thickness}).
As the strain increases from 2.3\% to 2.4\%, the change in the
average interlayer distance is $0.139 \Ang{}$, which is about 11 times
larger than the change in the average thickness of individual
\ce{MoSSe} layers. 
Thus the sharp decrease in the overall thickness is dominated by the
decrease in the interlayer distance. 
Differently, the gradual decrease in the overall thickness at small tensile
strain is dominated by the decrease in the thickness of individual
\ce{MoSSe} layers.

\begin{figure}[htb!]
\centering
\includegraphics[width=8.5cm]{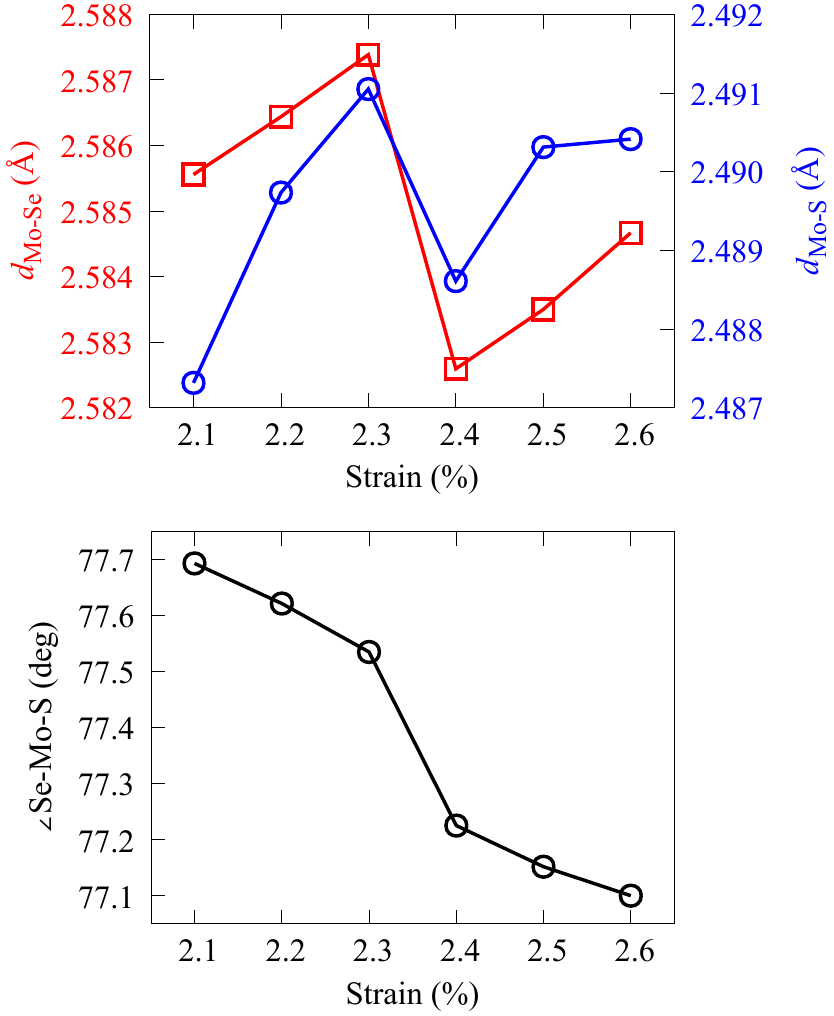} 
\caption{\label{fig:bonds} 
Bond lengths (upper panel) and bond angle (lower panel) of the \4th{}
\ce{MoSSe} layer of 4-layer \ce{MoSSe} near the critical strain.  
} 
\end{figure} 

In order to reveal the changes in bond length and bond angle within a
\ce{MoSSe} monolayer, we choose the \4th{} \ce{MoSSe} layer as a
representative. 
As shown in the upper panel of Fig.~\ref{fig:bonds}, 
both the \ce{Mo-Se} and \ce{Mo-S} bond lengths 
$d_\textrm{Mo-Se}$ and $d_\textrm{Mo-S}$ increase with strain 
before ($\textrm{strain}<2.3\%$) and after ($\textrm{strain}>2.4\%$) 
the sharp decrease in the overall thickness of
the 4-layer \ce{MoSSe}. 
However, $d_\textrm{Mo-Se}$ ($d_\textrm{Mo-S}$) decreases by about 
$ 0.005 \Ang $ ($ 0.003 \Ang $) as the strain increases from $2.3\%$ to $2.4\%$. 
The lower panel of Fig.~\ref{fig:bonds} shows the bond angle 
$\angle \ce{Se-Mo-S}$ versus the strain. 
The angle always decreases with strain, but the decrease 
is slightly faster for $2.3\% \leq \textrm{strain} \leq 2.4\%$. 
The reduction of the \ce{Mo-Se} and \ce{Mo-S} bond lengths together with
a faster decrease in the bond angle $\angle \ce{Se-Mo-S}$ signify the
sharp decrease of the thickness of a \ce{MoSSe} monolayer.

\section{Odd-even effect}
\label{sec:parity}

\setcounter{figure}{0}  

\begin{figure}[htb!]
\centering
\includegraphics[width=8.5cm]{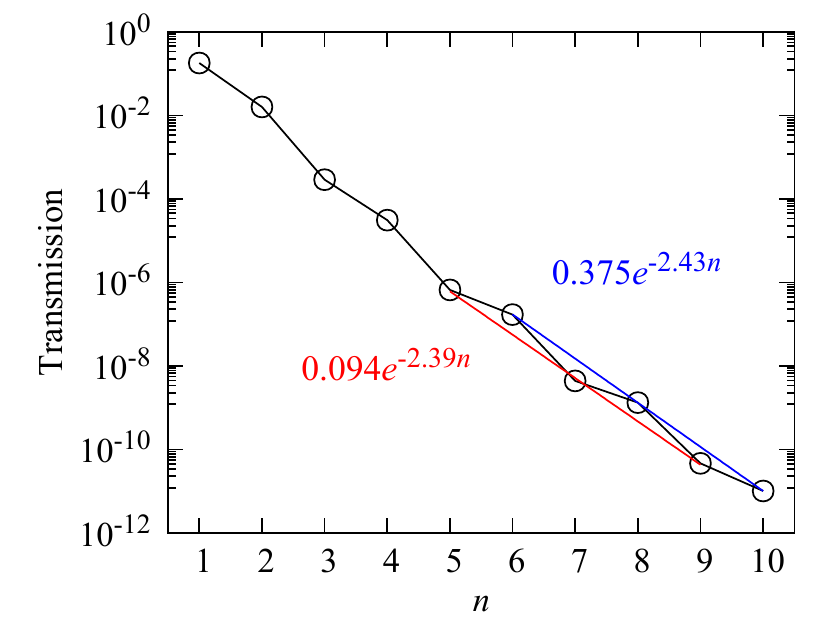} 
\caption{\label{fig:parity} 
Electron transmission at the Fermi energy for 
Zr/$n$-layer \ce{MoSSe}/Zr
junctions versus the number of %\ce{MoSSe} 
layers $n$. 
} 
\end{figure} 

\begin{figure}[htb!]
\centering
\includegraphics[width=8.5cm]{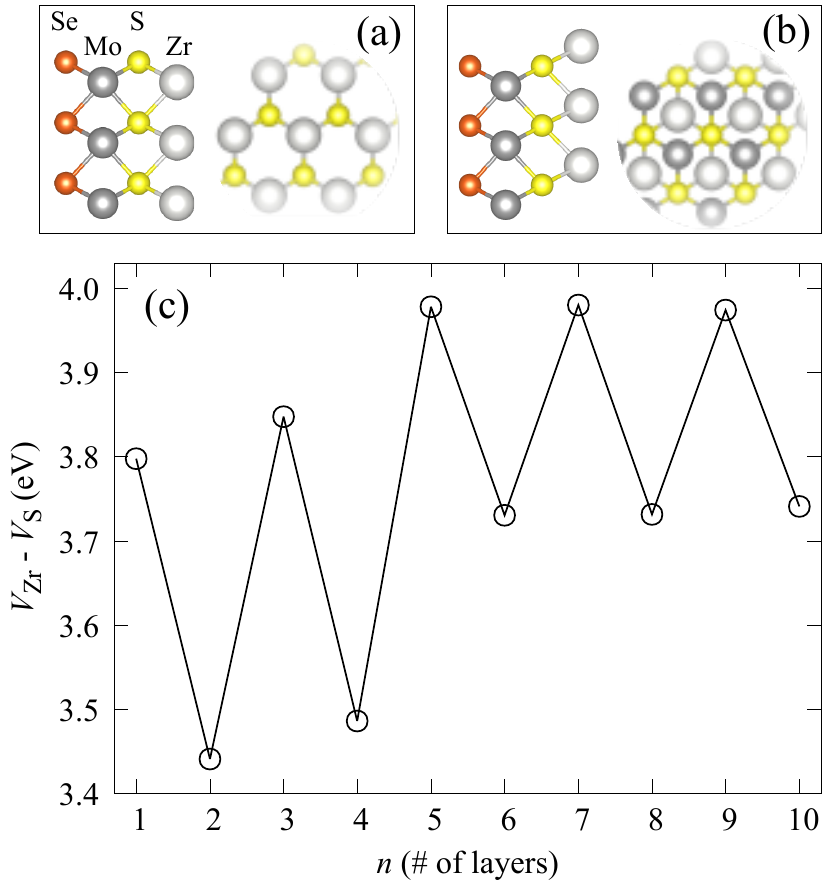} 
\caption{\label{fig:rightcontact} 
(a) [(b)] Top and side views of the contact between \ce{MoSSe} and Zr
on the right side of Zr/$n$-layer \ce{MoSSe}/Zr junctions with $n$ being
odd [even].
(c) Difference between the plane averaged electrostatic potential at Zr
and that at S at the contact shown in panels (a) and (b). 
} 
\end{figure} 

Fig.~\ref{fig:parity} shows the electron transmission at the Fermi energy $T_n(E_F)$ for Zr/$n$-layer \ce{MoSSe}/Zr junctions under zero bias
with $n=1, 2, 3, \dots, 10$. 
For simplicity, we will write $T_n(E_F)$ as $T_n$ in the rest of
this section. 
As seen from the figure, $T_n$ decays differently for even $n$ and
odd $n$. 
When $n \geq 5$, $T_n$ can be well fit by a function of the form
$T_n = \lambda e^{-\gamma n}$, 
$T^\textrm{o}_n \approx 9.44 \times 10^{-2} e^{-2.39}$
for odd $n$ and $T^\textrm{e}_n \approx 3.75 \times 10^{-1} e^{-2.43 n}$
for even $n$. 
The fitted curves are also shown in Fig.~\ref{fig:parity}. 
Interpolating, we see that $T^\textrm{o}_n < T^\textrm{e}_n$ at the
same number of layers $n$, or the same width of an effective tunneling
barrier. 
As such, we infer that the height of the effective tunneling barrier in odd
junctions is higher than in even junctions.

\begin{figure}[htb!]
\centering
\includegraphics[width=8.5cm]{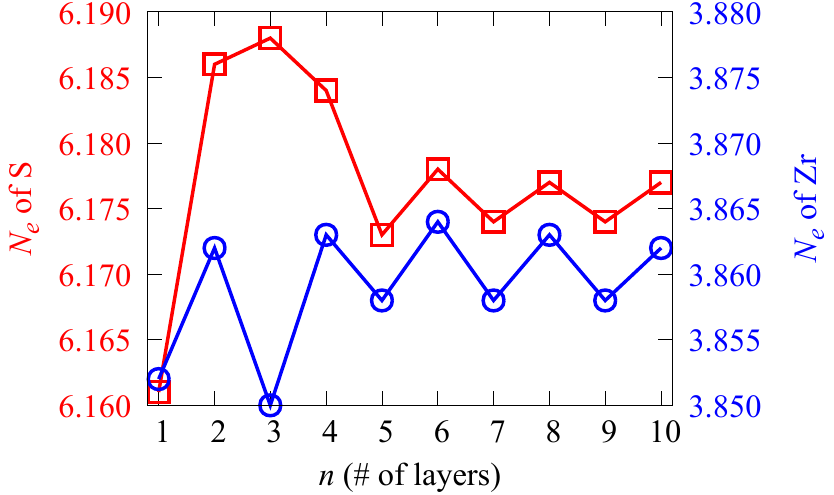} 
\caption{\label{fig:ne} 
Mulliken charge analysis: the number of electrons per S or Zr atom
at the right contact of Zr/$n$-layer \ce{MoSSe}/Zr junctions versus the
number of \ce{MoSSe} layers $n$. 
} 
\end{figure}

There are two possible reasons for such a difference in the height of the
tunneling barrier. 
The first one lies in the $n$-layer \ce{MoSSe} itself. 
Recall that the potential buildup at each van der Waals gap is
stacking-dependent and that the potential buildup at an \textit{h1} stacking
is larger. 
Given that the number of \textit{h1} stackings is greater (lesser) than the
number of \textit{h2} stackings by one in even (odd) junctions, we expect
that even junctions have a higher effective barrier. 
However, this is contradictory to the fact that $T^\textrm{o}_n < T^\textrm{e}_n$. 
The second possible reason pertains to the contact between the $n$-layer
\ce{MoSSe} and the Zr electrodes. 
The contact between the left Zr electrode and the $n$-layer \ce{MoSSe} is the same for all the junctions, 
while the contact on the right side is 
different for even and odd junctions, 
as depicted in Figs.~\ref{fig:rightcontact}a and \ref{fig:rightcontact}b. 
As a consequence, there is a parity-dependent potential difference between
the adjacent Zr and S atomic layers as shown in 
Fig.~\ref{fig:rightcontact}c. 
For $n \geq 5$, the odd junctions possess an $\sim 0.25 \eV{}$ higher potential
difference $V_\textrm{Zr} - V_\textrm{S}$ than the even junctions.
It seems that the potential buildup at the right contact competes with
the potential buildup at van der Waals gaps within a multilayer \ce{MoSSe}, 
with the former being dominant. 
Therefore, the effective tunneling barrier of odd junctions is higher than
that of even junctions. 
It is worth mentioning that the number of electrons at the right contact 
also exhibits an oscillatory behavior for $n \geq 5$, as shown in Fig. \ref{fig:ne}.

\bibliographystyle{apsrev4-2}
\bibliography{refs}

\end{document}